\documentclass[letterpaper,oneside,11pt]{article}

\usepackage[utf8]{inputenc}
\usepackage{amsthm}
\usepackage{amssymb}
\usepackage{amsmath,mathtools}
\usepackage{geometry}
\usepackage{commath}
\usepackage{graphicx}
\usepackage{subcaption}
\usepackage{bm}
\usepackage{float}
\usepackage{caption}
\usepackage{setspace}
\usepackage{multicol}
\usepackage{color}
\usepackage{subfiles}
\usepackage{textcomp}
\usepackage{enumerate}
\usepackage{hyperref}
\usepackage{commath}
\usepackage{mathrsfs}

\usepackage{comment}
\usepackage{natbib}
\usepackage{listings}
\usepackage{booktabs}
\usepackage[english]{babel}
\usepackage{multirow}
\usepackage{algorithm}
\usepackage[noend]{algpseudocode}
\usepackage{nicematrix, tikz-cd, tikz}
\usetikzlibrary{matrix}
\usetikzlibrary{shapes,decorations,decorations.pathreplacing,calligraphy,arrows,calc,arrows.meta,fit,positioning}
\tikzset{
    -Latex,auto,node distance =1 cm and 1 cm,semithick,
    state/.style ={ellipse, draw, minimum width = 0.7 cm},
    point/.style = {circle, draw, inner sep=0.04cm,fill,node contents={}},
    bidirected/.style={Latex-Latex,dashed},
    el/.style = {inner sep=2pt, align=left, sloped}
}  

\usepackage{setspace}
\usepackage[compact]{titlesec}
\titlespacing{\section}{0pt}{*0.8}{*0.8}
\titlespacing{\subsection}{0pt}{*0.8}{*0.8}
\titlespacing{\subsubsection}{0pt}{*0.8}{*0.8}

\newcommand{\bb}{ {\boldsymbol b} }

\newcommand{\bg}{ {\boldsymbol g} }

\newcommand{\bh}{ {\boldsymbol h} }

\newcommand{\bI}{ {\boldsymbol I} }
\newcommand{\bj}{ {\boldsymbol j} }

\newcommand{\bW}{ {\boldsymbol W} }
\newcommand{\bx}{ {\boldsymbol x} }

\newcommand{\by}{ {\boldsymbol y} }

\newcommand{\bz}{ {\boldsymbol z} }

\newcommand{\balpha}  { {\boldsymbol \alpha} }
\newcommand{\bbeta}   { {\boldsymbol \beta} }
\newcommand{\bgamma}  { {\boldsymbol \gamma} }

\newcommand{\bdelta}  { {\boldsymbol \delta} }

\newcommand{\bepsilon}{ {\boldsymbol \epsilon} }

\newcommand{\btheta}  { {\boldsymbol \theta} }

\newcommand{\bmu}     { {\boldsymbol \mu} }

\newcommand{\bsigma}  { {\boldsymbol \sigma} }

\newcommand{\bphi}    { {\boldsymbol \phi} }

\newcommand{\bzero}  { {\boldsymbol 0} }

\title{Integrative Variational Autoencoders for Generative Modeling of an Image Outcome with Multiple Input Images}
\author{Bowen Lei \and Yeseul Jeon
     \and Rajarshi Guhaniyogi
     \and 
     Aaron Scheffler \and Bani K. Mallick
   \and Alzheimer's Disease Neuroimaging Initiatives}
\date{\vspace{-5ex}}

\begin{document}

\maketitle
\begin{abstract}
Understanding relationships across multiple imaging modalities is a central goal in neuroimaging research. This work is motivated by the scientific challenge of predicting costly positron emission tomography (PET) scans using more accessible cortical structural measures derived from magnetic resonance imaging (MRI). We propose Integrative Variational Autoencoder (\texttt{InVA}), a novel and, to our knowledge, the \emph{first} hierarchical variational auto-encoder (VAE) framework designed for image-on-image regression in multimodal neuroimaging. \texttt{InVA} extends conventional VAEs to a predictive setting by modeling an outcome image as a function of both shared and modality-specific features from multiple input images. While standard VAEs are rarely applied to this type of regression task and are not designed to integrate information from multiple imaging sources, \texttt{InVA} effectively captures complex, nonlinear associations within and across images, while remaining computationally efficient. Unlike classical image-on-image regression methods that often rely on rigid model assumptions, \texttt{InVA} offers a highly flexible, model-free, data-driven alternative—crucial for modeling noisy neuroimaging data where such assumptions are difficult to justify. Empirical results demonstrate that \texttt{InVA} substantially outperforms conventional VAEs, as well as established nonlinear regression approaches such as Bayesian Additive Regression Trees (BART), which impose specific model constraints, and tensor regression methods, which cannot capture nonlinear dependencies. As a compelling application, \texttt{InVA} enables accurate  prediction of costly PET scans from cortical measures obtained through cost-effective structural MRI, offering a promising tool for integrative multimodal neuroimaging analysis.
\end{abstract}

\section{Introduction}\label{sec:Introduction}
A pressing challenge in contemporary neuroimaging research is to unravel the complex relationships among images capturing different facets of brain structure, with the goal of enabling accurate prediction of one imaging modality from others. This article shares a similar focus with motivation from a clinical application on patients suffering from Alzheimer's disease (AD), a neurodegenerative disorder characterized by progressive brain atrophy and cognitive decline. Central to the pathophysiological cascade that leads to AD is amyloid-$\beta$ (A$\beta$), a protein that accumulates into plaques in the brain of AD patients, and is thus a target for clinical therapeutics and molecular imaging \citep{hampel2021}. While PET with $\prescript{18}{}{\text{F-AV-45}}$ (florbetapir) radiotracer can characterize deposition of A$\beta$ in vivo to monitor disease progression and response to treatment, PET is a specialty imaging technique that is difficult to obtain and costly. It is of great interest to use more readily available MRI scans to reconstitute information from specialized and expensive A$\beta$ PET scans \citep{camus2012, zhang2022}. To this end, a natural approach would be to model A$\beta$ PET images from MRI derived metrics of cortical structure which have been shown to be associated with A$\beta$ deposition in patients with AD \citep{Spotorno2023}. Rather than considering a single measure of cortical structure, neuroscientists posit that multiple metrics (e.g. cortical thickness and volume) can be used as inputs to form a multi-modal imaging inputs which utilizes the cross-information among different images to improve prediction of A$\beta$ molecular images \citep{zhang2022, zhang2023}. To this end, Section~\ref{sec:image_on_image} offers a brief review of the existing literature on image-on-image regression in the context of predicting an output image from input images.

\subsection{Image-on-image Regression}\label{sec:image_on_image}
Image-on-image regression refers to the task of predicting one imaging modality using one or more other imaging modalities. This framework is especially valuable in scenarios where the target modality is either prohibitively expensive to acquire or when high-quality versions of the images are unavailable \citep{jeong2021restoration, subramanian2023complex, onishi2023self}.

A widely adopted strategy in this domain involves performing region-by-region regression between corresponding areas of the outcome and input images \citep{sweeney2013automatic}. While intuitive and computationally convenient, these region-wise approaches suffer from a major limitation: they fail to capture inter-regional dependencies, leading to reduced prediction accuracy. To partially address this limitation, methods such as pre-smoothing \citep{friston2003statistical} and adaptive smoothing \citep{qiu2007jump, yue2010adaptive} have been proposed to incorporate information from neighboring voxels. However, these smoothing techniques often fall short of capturing the complex spatial dependencies across regions and are limited in their ability to account for subject-specific heterogeneity. A more flexible alternative lies in spatially varying coefficient models, which allow regression coefficients to vary over space and are particularly well-suited for modeling spatial relationships between input and outcome images \citep{zhu2014spatially, mu2018estimation, mu2019spatially, niyogi2023tensor, guhaniyogi2022distributed, guhaniyogi2023distributed}. Building on this direction, spatial latent factor models have been introduced to model nonlinear and higher-order spatial dependencies \citep{guo2022spatial}. Despite their expressiveness, these models tend to be computationally intensive—even with moderate sample sizes and moderate number of brain regions—especially when attempting to capture the nonlinear structure inherent in brain imaging data.

Another promising line of work treats both input and outcome images as multi-dimensional arrays (tensors), giving rise to tensor-on-tensor regression models \citep{lock2018tensor, miranda2018tprm, guha2024covariate, guhaniyogi2020joint, guha2021bayesian, guhaniyogi2021bayesian}. These methods offer implicit spatial smoothing and leverage the tensor structure of imaging data. However, they often require downscaling of the images due to computational constraints and suffer from low signal-to-noise ratios. Moreover, they rely on restrictive linearity assumptions between input and outcome tensors, which may not capture the true complexity of the relationship between imaging modalities.
A third stream of research focuses on machine learning approaches, such as multivariate support vector machines, to predict missing spatial data in EEG using fMRI \citep{de2011predicting} or missing temporal data in fMRI from EEG \citep{jansen2012motion}. While powerful in certain contexts, these methods are often task-specific and may not generalize well across diverse imaging modalities or prediction settings.

Deep Neural Networks (DNNs) have become increasingly popular for image reconstruction tasks, thanks to their scalability with high-resolution images, large datasets, and capacity to model complex nonlinear relationships between input and outcome images. Among them, Convolutional Neural Networks (CNNs) are widely used in computer vision due to their ability to preserve spatial structures through convolutional layers. However, most CNN-based approaches are designed for task-specific applications \citep{santhanam2017generalized}, and commonly use architectures like Visual Geometry Group (VGG) networks and Residual Neural Networks (ResNet), which may inadvertently incorporate irrelevant features, leading to biased predictions \citep{isola2017image}. To support more general-purpose image-on-image regression, the Recursively Branched Deconvolutional Network (RBDN) was proposed. RBDN constructs a composite feature map that is processed through multiple task-specific convolutional branches \citep{santhanam2017generalized}. Despite its flexibility, RBDN requires input and outcome images to have identical dimensions, which limits its applicability when image sizes vary.

Compared to high-dimensional and complex images, carefully extracted low-dimensional representations can substantially improve the estimation of relationships between input and outcome images. To achieve this, recent work has leveraged deep generative models, particularly variational autoencoders (VAEs) \citep{kingma2013auto, goodfellow2014generative, rezende2014stochastic, li2015generative, doersch2016tutorial, girin2020dynamical, zhao2023revisiting}, which have shown notable success in image reconstruction tasks. VAEs introduce a latent variable, often modeled with a simple multivariate Gaussian distribution, to encode compressed data representations. The encoder maps an image into this latent space, and the decoder reconstructs the image from samples drawn from the latent representation. 
By capturing essential image features in a much lower-dimensional space, VAEs enable regression tasks to be performed efficiently within the latent space, jointly with encoder–decoder training.

A key strength of VAEs lies in their use of flexible probabilistic frameworks that capture salient image features, while remaining computationally scalable for high-dimensional inputs and large datasets. However, despite their success in single-input image modeling, most existing VAE-based approaches are not designed to effectively leverage shared information across multiple input images when predicting an outcome image. Specifically, VAE strategies with multiple imaging inputs typically rely on either \emph{input-level fusion}, where multiple input images are concatenated prior to modeling (\citep{ren2021infrared,duffhauss2022fusionvae})—or \emph{decision-level fusion}, where separate models are trained for each modality and their outputs are later combined (\citep{kurle2019multi,du2021two}). Both strategies have limitations: input-level fusion can lead to excessively large feature spaces and requires careful design choices about how inputs are merged, while decision-level fusion ignores synergistic and complementary information across modalities during training. Recent studies in multimodal neuroimaging provide strong evidence that joint modeling of shared information across images significantly improves prediction accuracy compared to separate or naively combined inputs \citep{gutierrez2024multi, guha2024bayesian, jeon2025deep}. Nevertheless, this remains an underexplored area in the VAE literature, especially for image-on-image regression with multiple input images.

\subsection{Our Contributions}

In multi-modal neuroimaging, hierarchical Bayesian methods offer a principled approach to borrowing structured information across imaging inputs by imposing joint priors on model parameters at different levels of the hierarchy. This facilitates coherent inference via the joint posterior distribution \citep{jin2020bayesian, su2022comparative, kaplan2023bayesian}. However, despite their theoretical appeal, such approaches remain underutilized due to significant computational challenges and the absence of scalable modeling architectures. 

Motivated by the hierarchical Bayesian principle of leveraging shared structure across data sources, this paper presents the Integrative Variational Autoencoder (InVA)—a novel and computationally efficient framework for predicting imaging outcome from multiple imaging inputs. InVA operates in two interconnected stages. In the first stage, it constructs image-specific deep neural network (DNN) encoders and decoders for each of the  input images, enabling each input image to be mapped into its own low-dimensional latent space. This design preserves flexibility in modeling the unique features of each image, providing a representation referred to as \emph{shallow features}. Concurrently, a shared encoder–decoder pair transforms the \emph{shallow features} into \emph{deep features} that encode cross-image dependencies and shared latent structure. In the second stage, summaries of shared and image-specific encoding distributions are jointly fed into a DNN-based prediction network tasked with reconstructing the outcome image. The image-specific encoder-decoder components, shared encoder-decoder components, and predictive network are trained \emph{jointly}, ensuring that feature learning, reconstruction, and prediction mutually reinforce one another. This unified optimization strategy enables InVA to disentangle complementary image-specific and shared information, yielding coherent latent representations and robust outcome prediction.

Empirical evaluations demonstrate that \texttt{InVA} consistently outperforms conventional VAEs trained separately on individual input images, as well as other leading image-on-image regression approaches. Its hierarchical design and joint optimization strategy allow it to seamlessly integrate diverse imaging inputs, resulting in significantly enhanced predictive performance.

\subsection{Innovation Over Hierarchical Variational Auto-Encoder Literature}
Our proposed approach introduces a novel hierarchical modeling architecture that goes well beyond the traditional goals of hierarchical VAEs. 
Notably, prior work on hierarchical VAEs has focused primarily on enhancing the expressiveness of generative models. For example, in the hierarchical VAE literature, DRAW~\citep{gregor2015draw} introduces a sequential, attention-based VAE for more realistic image generation using a recurrent encoder–decoder framework. Ladder VAE~\citep{sonderby2016ladder} improves generative accuracy by recursively correcting the latent distribution across layers. This is further generalized to other hierarchical variational models to get expressive variational distribution as well as efficient computation~\citep{ranganath2016hierarchical}. Hierarchical priors proposed in \cite{klushyn2019learning} aim to overcome over-regularization from standard normal priors on latent representations in VAEs by incorporating more structured prior distributions to induce useful latent representations. More recently, NVAE~\citep{vahdat2020nvae} designed a hierarchical VAE, which utilizes a deep hierarchical structure to achieve more stable and accurate image reconstruction. 

While these contributions have greatly improved the quality of unsupervised image reconstruction, they are not suited for the supervised prediction of an output image from multiple input images. They typically do not leverage cross-image relationships or jointly model shared and image-specific latent structure needed for image-on-image regression. By contrast, InVA is the first hierarchical VAE designed explicitly to integrate multiple imaging inputs for outcome prediction. Its key novelties are the following. \textbf{(1) Supervised Predictive Framework:} Unlike conventional hierarchical VAEs focused on generative modeling, InVA is tailored for image-on-image regression, directly linking input images to an outcome image using supervised training objectives.
\textbf{(2) Joint Modeling of Shared and Image-Specific Representations:} InVA explicitly constructs both image-specific (shallow) and shared (deep) latent representations through image-specific and shared encoders/decoders. This layered design enables it to capture both unique and common information across images—a capability absent in existing hierarchical VAE literature. \textbf{(3) Hierarchical Feature Fusion for Prediction:} Rather than using hierarchical structure solely to improve variational approximations, InVA uses it to fuse complementary features across inputs to inform prediction. This design leads to enhanced flexibility, accuracy, and interpretability. \textbf{(4) Scalability and Efficiency:} InVA is computationally efficient and scalable, avoiding the expensive inference schemes used in many deep hierarchical VAEs, making it well-suited for neuroimaging applications with a large number of subjects. \textbf{(5) Enhanced Predictive Accuracy:} Empirical results demonstrate that InVA significantly outperforms traditional VAEs (even hierarchical ones) and other state-of-the-art image-on-image regression methods, due to its integrative and supervised learning design.

\section{Proposed Approach}
We propose an Integrative Variational Autoencoder (\texttt{InVA}) to better integrate multiple imaging inputs for more accurate prediction of an imaging output. We first begin by defining notations and offering a brief overview on VAEs.
\subsection{Notations}
For $i=1,...,n$, we observe $K$ different imaging inputs $\bx_{1,i},...,\bx_{K,i}$ from the $i$th subject, with $\bx_{k,i}\in\mathbb{R}^{J_k}$, $k=1,...,K$, and the corresponding outcome image $\by_i\in\mathbb{R}^m$. We denote the input data for the $i$th subject to be $\bx_{(i)}=\{\bx_{1,i},...,\bx_{K,i}\}$.


\begin{figure*}[!t]
    \centering
    \includegraphics[width = 1.0\textwidth]{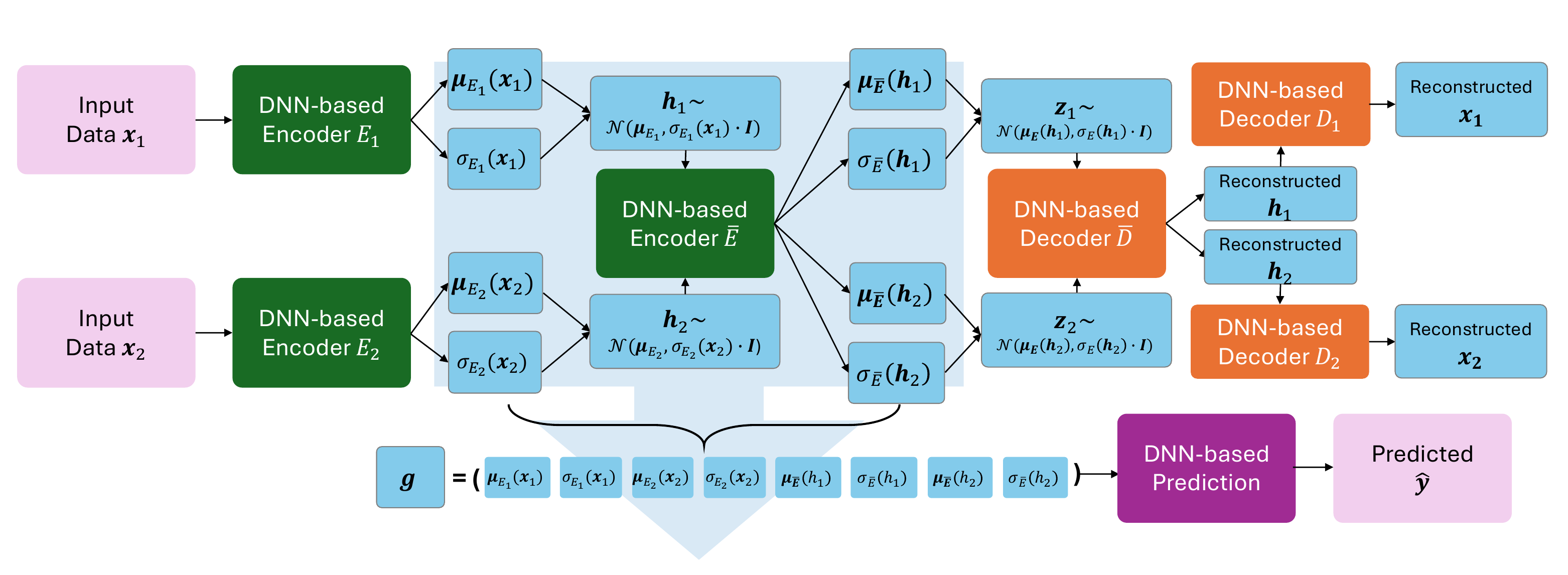}
    \caption{Architecture of Integrative Variational Autoencoder  (\texttt{InVA}), which includes modality-specific encoding networks $E_k$, $k\in \{1,\cdots,K\}$ (in green), shared encoding network $\bar{E}$ (in green), shared decoding network $\bar{D}$ (in orange), and image-specific decoding network $D_k$, $k\in \{1,\cdots,K\}$ (in orange). It also shows the prediction of the output image $\hat{\by}$ based on the concatenated
feature vector of means and standard deviations for shallow and deep feature vectors for shared
and image-specific autoencoders. For the purpose of illustration, we show the architecture for $K=2$.}\label{fig:hvae}
\end{figure*}


\subsection{Preliminary: Variational Autoencoder}

Autoencoder (AE) is a widely-used unsupervised learning method that utilizes an encoder to compress data and reconstruct the data from the encoded features through a decoder~\citep{geng2015high, tschannen2018recent, chorowski2019unsupervised, nazari2023geometric, hao2023coupled}. To cope with different scenarios, variants of autoencoders have also been inspired~\citep{ng2011cs294a, rifai2011higher, rifai2011contractive, chen2012marginalized, chen2014marginalized, ranjan2017hyperface, kingma2013auto, tolstikhin2017wasserstein, pei2018study, vahdat2020nvae}.
Based on AE, variational autoencoders (VAEs) are designed to model the data distribution~\citep{doersch2016tutorial, girin2020dynamical, zhao2023revisiting}, which maps the input data into latent Gaussian distribution through the encoder~\citep{kviman2023cooperation, hao2023coupled, janjos2023unscented}.

The standard Variational Autoencoder (VAE) framework \citep{kingma2013auto} consists of two primary components: an encoder and a decoder. The encoder, denoted as $q_\bphi(\bz_i|\bx_i)$, maps the input 
$\bx_i\in\mathbb{R}^J$ to a latent representation $\bz_i\in\mathbb{R}^p$. This encoder is typically modeled as a multivariate Gaussian: $q_\bphi(\bz_i|\bx_i)=N(\bz_i|\bmu_E(\bx_i;\bphi),\bsigma_E(\bx_i;\bphi)^2),$ where the $p$-variate functions $\bmu_E(\bx_i;\bphi)$ and $\bsigma_E(\bx_i;\bphi)$ define the mean and standard deviation, respectively. These functions are parameterized jointly using a fully connected deep neural network (DNN), with $\bphi$ representing the corresponding weights and biases. 

The prior distribution on latent variables are typically assumed to be standard normal, i.e., $p(\bz_i)=N(\bzero,\bI_p)$. The decoder, denoted by $p_\btheta(\bx_i|\bz_i)$, reconstructs the input $\bx_i$
from its latent encoding $\bz_i$, and is generally modeled as a multivariate normal distribution with identity covariance:
$p_\btheta(\bx_i|\bz_i)=N(\bx_i|\bmu_D(\bz_i;\btheta),\bI_J)$, 
where $\bmu_D(\bz_i;\btheta)\in\mathbb{R}^J$ is learned via another fully connected DNN with $\btheta$ as the parameters. The reparameterization trick \citep{blei2017variational} is employed to enable gradient-based optimization through backpropagation.

The overall objective of the VAE is to approximate the marginal likelihood $p(\bx_1,...\bx_n)=\int p(\bx_1,...,\bx_n|\bz_1,...,\bz_n)p(\bz_1,..,\bz_n)d\bz_1\cdots d\bz_n$, which is generally intractable. To circumvent this, an amortized inference strategy is used by introducing a variational posterior of the form $q_\bphi(\bz_1,...,\bz_n|\bx_1,..,\bx_n)=\prod_{i=1}^n q_\bphi(\bz_i|\bx_i)$.
The model is trained by maximizing the evidence lower bound (ELBO) on the log-marginal likelihood:
\begin{align*}
\log p(\bx_1, \dots, \bx_n) &= E_{q_\bphi} \left[ \log p(\bx_1, \dots, \bx_n, \bz_1, \dots, \bz_n) - \log q_\bphi(\bz_1, \dots, \bz_n | \bx_1, \dots, \bx_n) \right] \\
&\quad + \text{KL}\left( q_\bphi(\bz_1, \dots, \bz_n | \bx_1, \dots, \bx_n) | p(\bz_1, \dots, \bz_n | \bx_1, \dots, \bx_n) \right) \\
&\geq E_{q_\bphi} \left[ \log p(\bx_1, \dots, \bx_n, \bz_1, \dots, \bz_n) - \log q_\bphi(\bz_1, \dots, \bz_n | \bx_1, \dots, \bx_n) \right].\\
&=\sum_{i=1}^n \left\{E_{q_\bphi}[\log p_{\btheta}(\bx_i|\bz_i)]-\text{KL}(q_{\bphi}(\bz_i|\bx_i)||p(\bz_i))\right\}.
\end{align*}
This leads to the following ELBO expression
\begin{align}
    \text{ELBO}(\btheta,\bphi) &= \sum_{i=1}^n \bigg{\{} 
    E_{q_\bphi}\big{[}\log p_{\btheta}(\bx_{i}|\bz_{i})-\text{KL} \big{(} q_\bphi(\bz_{i}|\bx_{i}) || p(\bz_{i}) \big{)}\big{]}  \bigg{\}}\nonumber\\
    &= \sum_{i=1}^n \bigg{\{} -||\bx_{i} - \hat{\bx}_{i}||_2^2+ \frac{1}{2}\sum_{j=1}^p( \log \sigma_{E,j}(\bx_i;\bphi)^2 - \mu_{E,j}(\bx_i;\bphi)^2 - \sigma_{E,j}(\bx_i;\bphi)^2 + 1 ) \bigg{\}},\label{eq:vae-loss-2}
\end{align}
where $\hat{\bx}_i=\bmu_D(\bz_i;\btheta)$ is the reconstructed input, and $\mu_{E,j}(\bx_i;\bphi)$ and $\sigma_{E,j}(\bx_i;\bphi)$ are the $j$th elements of $\bmu_{E}(\bx_i;\bphi)$ and $\bsigma_{E}(\bx_i;\bphi)$, respectively.

While the VAE formulation above is developed for unsupervised inference, our focus lies on the supervised prediction of an outcome $\by_i$ using both shared and image-specific information from the input images $\bx_{1,i},\ldots,\bx_{k,i}$. This supervised setting is largely unexplored in the literature, which we address in the next section.

\subsection{Integrative Variational Autoencoder}
We propose an architecture inspired by hierarchical Bayesian modeling to improve the learning of latent variable distributions from multiple imaging inputs. This integrative variational autoencoder (In-VA) is designed to capture both image-specific and shared structures in a principled way. At a shallow level, the architecture includes separate encoders and decoders for each input image to extract and reconstruct features unique to that image. At a deeper level, it incorporates encoders and decoders shared by all images to promote information borrowing and capture common patterns present across different inputs. This hierarchical design mirrors the structure of multi-level Bayesian models, where image-specific parameters capture within-image variation while shared parameters capture between-image dependence.

\noindent \textbf{Image-specific encoder:} 
For each input image $\mathbf{x}_{k,i}\in\mathbb{R}^{J_k}$, the image-specific encoder, denoted as $q_{\balpha_k}(\bh_{k,i}|\bx_{k,i})$, maps the image into a latent representation $\bh_{k,i}\in\mathbb{R}^{p}$, referred to as \emph{shallow features}. This encoder is modeled as a multivariate Gaussian:
$q_{\balpha_k}(\bh_{k,i}|\bx_{k,i})=N(\bh_{k,i}|\bmu_{E_k}(\bx_{k,i};\balpha_k),\sigma_{E_k}(\bx_{k,i};\balpha_k)^2\bI_{p})$, where the mean and variance functions $\bmu_{E_k}(\bx_{k,i};\balpha_k)\in\mathbb{R}^{p}$ and $\sigma_{E_k}(\bx_{k,i};\balpha_k)\in\mathbb{R}$ are jointly modeled using a deep neural network architecture given by the following: 
\begin{equation}
    (\bmu_{E_k}(\bx_{k,i};\balpha_k)^T, \log\sigma_{E_k}(\bx_{k,i};\balpha_k))^T = \sigma_{L} \left( \mathbf{W}_{k,L}^{(E)} \sigma_{L-1} \left( \cdots \sigma_{2} \left( \mathbf{W}_{k,1}^{(E)} \mathbf{x}_{k,i} + \mathbf{b}_{k,1}^{(E)} \right) \cdots + \mathbf{b}_{k,L-1}^{(E)} \right) + \mathbf{b}_{k,L}^{(E)} \right),
\label{eq:isencoder}\end{equation}
where the weight matrix $\bW_{k,l}^{(E)}\in\mathbb{R}^{o_{k,l}^{(E)}\times o_{k,l-1}^{(E)}}$ connects the $o_{k,l-1}^{(E)}$ neurons in the $(l-1)$th layer to the $o_{k,l}^{(E)}$ neurons of the $l$th hidden layer, $\sigma_{l}(\cdot)$ is the activation function for the $l$th layer and $\mathbf{b}_{k,l}\in\mathbb{R}^{o_{k,l}}$ corresponds to the bias parameter at the $l$th layer. The number of neurons at the $l$th layer is given by $o_{k,l}^{(E)}$. We loosely refer to this image-specific encoding network as $E_k$. The weights and bias parameters for the deep neural network, denoted collectively by $\balpha_k=\{\bW^{(E)}_{k,1},...,\bW^{(E)}_{k,L},\bb^{(E)}_{k,1},...,\bb^{(E)}_{k,L}\}$, determine how raw input images are transformed into low-dimensional shallow features capturing image-specific information. 

\noindent \textbf{Shared encoder:} While the image-specific encoder focuses on unique features of each image, the shared encoder $q_{\bbeta}(\bz_{k,i}|\bh_{k,i})$ maps the \emph{shallow features} $\bh_{k,i}\in\mathbb{R}^{p}$ into \emph{deep features} $\bz_{k,i}\in\mathbb{R}^{q}$ that capture relationships common across all modalities. Like the image-specific encoder, the shared encoder is modeled as a multivariate Gaussian distribution, $q_{\bbeta}(\bz_{k,i}|\bh_{k,i})=N(\bz_{k,i}|\bmu_{\bar{E}}(\bh_{k,i};\bbeta),\sigma_{\bar{E}}(\bh_{k,i};\bbeta)^2\bI_q)$, where the functions $\bmu_{\bar{E}}(\bh_{k,i};\bbeta)\in\mathbb{R}^q$ and $\sigma_{\bar{E}}(\bh_{k,i};\bbeta)\in\mathbb{R}$ are jointly modeled using a deep neural network architecture given by the following: 
\begin{equation}
    (\bmu_{\bar{E}}(\bh_{k,i};\bbeta)^T, \log\sigma_{\bar{E}}(\bh_{k,i};\bbeta))^T = \sigma_{L} \left( \mathbf{W}_{L}^{(\bar{E})} \sigma_{L-1} \left( \cdots \sigma_{2} \left( \mathbf{W}_{1}^{(\bar{E})} \mathbf{h}_{k,i} + \mathbf{b}_{1}^{(\bar{E})} \right) \cdots + \mathbf{b}_{L-1}^{(\bar{E})} \right) + \mathbf{b}_{L}^{(\bar{E})} \right),
\label{eq:ishencoder}\end{equation}
where the weight matrix $\bW_{l}^{(\bar{E})}\in\mathbb{R}^{o_{l}^{(\bar{E})}\times o_{l-1}^{(\bar{E})}}$ connects the $o_{l-1}^{(\bar{E})}$ neurons at the $(l-1)$th layer to the $o_{l}^{(\bar{E})}$ neurons at the $l$th hidden layer, $\sigma_{l}(\cdot)$ is the activation function for the $l$th layer and $\mathbf{b}_{l}^{(\bar{E})}\in\mathbb{R}^{o_{l}^{(\bar{E})}}$ corresponds to the bias parameter at the $l$th layer. The
shared encoder parameter $\bbeta=\{\bW_{1}^{(\bar{E})},...,\bW_{L}^{(\bar{E})},\bb_{1}^{(\bar{E})},...,\bb_{L}^{(\bar{E})}\}$ are common to all images, ensuring that the learned deep features reside in a unified latent space where cross-image patterns can be modeled effectively. We loosely refer to this shared encoding network as $\bar{E}$.

\noindent\textbf{Shared decoder:} The shared decoder $p_{\bgamma}(\bh_{k,i}|\bz_{k,i})$ performs the inverse mapping of the shared encoder, reconstructing the shallow features $\bh_{k,i}$ from deep features $\bz_{k,i}$. This mapping is modeled with a multivariate normal distribution, given by,
$p_{\bgamma}(\bh_{k,i}|\bz_{k,i})=N(\bh_{k,i}|\bmu_{\bar{D}}(\bz_{k,i};\bgamma),\bI_p)$, where $\bmu_{\bar{D}}(\bz_{k,i};\bgamma)\in\mathbb{R}^{p}$ is an unknown function modeled using a deep neural network architecture given by,
\begin{equation}
    \bmu_{\bar{D}}(\bz_{k,i};\bgamma)= \sigma_{L} \left( \mathbf{W}_{L}^{(\bar{D})} \sigma_{L-1} \left( \cdots \sigma_{2} \left( \mathbf{W}_{1}^{(\bar{D})} \mathbf{z}_{k,i} + \mathbf{b}_{1}^{(\bar{D})} \right) \cdots + \mathbf{b}_{L-1}^{(\bar{D})} \right) + \mathbf{b}_{L}^{(\bar{D})} \right).
\label{eq:ishdecoder}\end{equation}
Here $\bW_{l}^{(\bar{D})}\in\mathbb{R}^{o_{l}^{(\bar{D})}\times o_{l-1}^{(\bar{D})}}$ is the weight matrix and $\bb_l^{(\bar{D})}\in\mathbb{R}^{o_{l}^{(\bar{D})}}$ is the bias vector. The parameter $\bgamma$ represents the set of all weight and bias parameters $\{\mathbf{W}_{1}^{(\bar{D})},...,\mathbf{W}_{L}^{(\bar{D})},\bb_{1}^{(\bar{D})},...,\bb_{L}^{(\bar{D})}\}$ that ensures that the deep features can be transformed back into shallow features before final reconstruction. The shared decoding network is loosely referred to as $\bar{D}$.

\noindent \textbf{Image-specific decoder:} Once the shared decoder has reconstructed the shallow features $\bh_{k,i}$, an image-specific decoder $p_{\btheta_k}(\bx_{k,i}|\bh_{k,i})$ maps these features back to the original $k$th input image space. This decoder is modeled as a multivariate Gaussian with mean function $\bmu_{D_k}(\bh_{k,i};\btheta_k)\in\mathbb{R}^{J_k}$ and an identity covariance matrix, given by, $p_{\btheta_k}(\bx_{k,i}|\bh_{k,i})=N(\bx_{k,i}|\mu_{D_k}(\bh_{k,i};\btheta_k),\bI_{J_k})$. Similar to the shared-decoder, the mean function $\mu_{D_k}(\bh_{k,i};\btheta_k)$ for image-specific decoders are modeled as a deep neural network, 
\begin{equation}
    \bmu_{D_k}(\bh_{k,i};\btheta_k)= \sigma_{L} \left( \mathbf{W}_{L,k}^{(D)} \sigma_{L-1} \left( \cdots \sigma_{2} \left( \mathbf{W}_{1,k}^{(D)} \mathbf{h}_{k,i} + \mathbf{b}_{1,k}^{(D)} \right) \cdots + \mathbf{b}_{L-1,k}^{(D)} \right) + \mathbf{b}_{L,k}^{(D)} \right).
\label{eq:isdecoder}\end{equation}
Here $\bW_{l,k}^{(D)}\in\mathbb{R}^{o_{l,k}^{(D)}\times o_{l-1,k}^{(D)}}$ is the weight matrix and $\bb_{l,k}^{(D)}\in\mathbb{R}^{o_{l,k}^{(D)}}$ is the bias vector. The parameter $\btheta_k$ represents the set of all weight and bias parameters $\{\mathbf{W}_{1,k}^{(D)},...,\mathbf{W}_{L,k}^{(D)},\bb_{1,k}^{(D)},...,\bb_{L,k}^{(D)}\}$.
The image-specific decoder is responsible for capturing the fine-grained structural and intensity details unique to each image, enabling high-fidelity reconstruction. The image-specific decoding network is loosely referred to as $D_k$.

Overall, this hierarchical encoder–decoder architecture allows the \texttt{InVA} to disentangle image-specific variation from multi-image input structure, leading to latent representations that are both rich in individual image detail and coherent across images. The two-stage decoding process—shared decoding to recover shallow features followed by image-specific decoding to reconstruct the original image—ensures that both shared and unique characteristics of each input are preserved in the generative process. The flowchart illustrating the model development process, incorporating both shared and image-specific encoders and decoders, is presented in  Figure~\ref{fig:hvae}.

\noindent\textbf{Deep neural network-based prediction of the output image:} 
As illustrated in Figure \ref{fig:hvae}, the final stage of our framework predicts the target output image $\by_i\in\mathbb{R}^m$ using deep neural network (DNN) predictor layers. The input to this prediction module is a concatenated feature vector of means and standard deviations for shallow and deep feature vectors for shared and image-specific autoencoders, given by,
\begin{align}
\bg_i=(\bmu_{E_k}(\bx_{i,k};\balpha_k)^T,\sigma_{E_k}(\bx_{i,k};\balpha_k),\bmu_{\bar{E}}(\bh_{i,k};\bbeta)^T, \sigma_{\bar{E}}(\bh_{i,k};\bbeta):k=1,..,K).   
\end{align}
This feature vector encapsulates both image-specific and shared latent representations, thereby providing a comprehensive summary of the input information. 
The predictor network maps $\bg_i$ to the predicted output image $\widehat{y}_i$ through a sequence of $L$ fully connected layers with nonlinear activations:
\begin{equation}\label{predDNNn}
    \widehat{y}_i = \sigma_{L} \left( \mathbf{W}_L^{(y)} \sigma_{L-1} \left( \cdots \sigma_2 \left( \mathbf{W}_{1}^{(y)}\bg_i + \mathbf{b}_1^{(y)} \right) \cdots + \mathbf{b}_{L-1}^{(y)} \right) + \mathbf{b}_{L}^{(y)} \right),
\end{equation}
where $\mathbf{W}_l^{(y)}$ denotes the weight matrix connecting the $(l-1)$th layer and $l$th layer, and $\mathbf{b}_l^{(y)}$ denotes the bias vector $l$th layer, respectively, and $\sigma_l(\cdot)$ represents the activation function applied at that layer. Here $\bdelta$ denotes the set of all weight and bias parameters for the deep neural network specified in Equation~\eqref{predDNNn}. By learning a flexible nonlinear mapping from the fused latent representation to the output space, this DNN-based predictor is able to exploit complex interdependencies among the multi-modal features, allowing for accurate and high-fidelity output image prediction. 

\subsection{Model Training}
Training the proposed framework is designed to jointly optimize three objectives: (a) \emph{reconstruction fidelity:} accurately reconstructing the input images from their latent representations; (b) \emph{latent space regularization:} enforcing structured, well-behaved latent variables via variational inference; (c) \emph{prediction accuracy:} producing accurate estimates of the target output image.
To achieve this, we formulate a joint loss function composed of two complementary components:
\begin{itemize}
\item \textbf{Reconstruction loss:} This is denoted as $\mathcal{L}_{reconstruction}$, which measures the ability of the integrative variational autoencoder (InVA) to reconstruct each input image from its latent features.
\item \textbf{Prediction loss:} This is denoted as $\mathcal{L}_{prediction}$, which measures the accuracy of predicting the output image $\by_i$ from the learned latent representations.
\end{itemize}
The total training loss is then expressed as:
\begin{align}\label{eq:total_loss}
 \mathcal{L}_{total}(\{\balpha_k,\btheta_k:k=1,..,K\},\bbeta,\bgamma,\bdelta)=  \mathcal{L}_{reconstruction}(\{\balpha_k,\btheta_k:k=1,..,K\},\bbeta,\bgamma)+  \mathcal{L}_{prediction}(\bdelta). 
\end{align}
Importantly, the reconstruction loss from the \texttt{InVA} framework and the prediction loss for the output image are not optimized in isolation. Instead, they are simultaneously minimized, enabling effective information sharing between the unsupervised representation learning of the input images and supervised prediction of the output image. This joint training ensures that fine-grained imaging details captured during reconstruction can inform the prediction model, while predictive supervision helps the encoder focus on features that are most relevant for downstream tasks, rather than preserving irrelevant variation in the inputs. The result is a model that borrows strength across images, tasks, and representation levels, leading to more robust latent features and better overall performance compared to training the reconstruction and prediction components separately. We offer a description of the two loss functions below.

\noindent \textbf{Loss function for input image reconstruction:} In constructing the shallow and deep latent features, our goal is to maximize the marginal likelihood of the input images across all modalities. Notably, the marginal likelihood of the input images can be expressed as:
\begin{align}\label{eq:vi-1}
&\log p(\{\bx_{k,i}:k=1,..,K;\:i=1,..,n\})=\sum_{i=1}^n\log p(\{\bx_{k,i}:k=1,..,K\})\nonumber\\
&=\sum_{i=1}^n \Bigg[KL(q_{\balpha_k}(\bh_{(i)}|\bx_{(i)})||p(\bh_{(i)}|\bx_{(i)}))+KL(q_{\bbeta}(\bz_{(i)}|\bh_{(i)})||p(\bz_{(i)}|\bh_{(i)}))+\nonumber\\
&\qquad\quad\mathcal{L}(\{\balpha_k,\btheta_k:k=1,..,K\},\bbeta,\bgamma,\bx_{(i)})\Bigg].
\end{align}
In Equation~(\ref{eq:vi-1}), The first KL term comes from the image-specific encoder, and it penalizes deviations between the approximate posterior distribution of the shallow features $\bh_{(i)}=\{\bh_{1,i},...,\bh_{K,i}\}$ and their prior. The second KL term comes from the shared encoder, and it penalizes deviations between the approximate posterior distribution of the deep features $\bz_{(i)}=\{\bz_{1,i},...,\bz_{K,i}\}$ and their prior. Both KL terms act as latent regularizers, encouraging the learned latent distributions to remain close to predefined priors (isotropic Gaussian distributions), which prevents overfitting and promotes generalization. The third term $\mathcal{L}(\{\balpha_k,\btheta_k:k=1,..,K\},\bbeta,\bgamma)$ is the ELBO term which can be written as:
\begin{align}
    &\mathcal{L}(\{\balpha_k,\btheta_k:k=1,..,K\},\bbeta,\bgamma, \bx_{(i)}) = \sum_{k=1}^K \bigg{\{} E_{q_{\balpha_{k}}(\bh_{k,i}|\bx_{k,i}) q_{\bbeta}(\bz_{k,i}|\bh_{k,i})} [\log p_{\btheta_k,\bgamma}(\bx_{k,i}|\bz_{k,i})] - \nonumber \\
    &\qquad\qquad\qquad\text{KL}\big{(}  q_{\balpha_{k}}(\bh_{k,i}|\bx_{k,i}) | p(\bh_{k,i}) \big{)} - \text{KL}\big{(} q_{\bbeta}(\bz_{k,i}|\bh_{k,i}) | p(\bz_{k,i}) \big{)} \bigg{\}},\label{eq:hvae-vi}
\end{align}
where $p(\bz_{k,i})$ and $p(\bh_{k,i})$ are prior distributions on the deep and shallow features, respectively. Both prior distributions are taken to be multivariate normal with zero mean and covariance as the identity matrix.
Given that the first two terms in Equation~\eqref{eq:vi-1} are nonnegative, maximizing the marginal likelihood is equivalent to maximizing the ELBO term in Equation~\eqref{eq:hvae-vi}. Hence, the loss function due to reconstruction of input images is defined as the negative of the ELBO term given by 
\begin{align}
&\mathcal{L}_{reconstruction}(\{\balpha_k,\btheta_k:k=1,..,K\},\bbeta,\bgamma)= - \sum_{i=1}^n\mathcal{L}(\{\balpha_k,\btheta_k:k=1,..,K\},\bbeta,\bgamma, \bx_{(i)})\nonumber\\
&= \sum_{i=1}^n\sum_{k=1}^K \bigg{\{} \text{KL}\big{(}  q_{\balpha_{k}}(\bh_{k,i}|\bx_{k,i}) | p(\bh_{k,i}) \big{)} + \text{KL}\big{(} q_{\bbeta}(\bz_{k,i}|\bh_{k,i}) | p(\bz_{k,i})\big{)}-\nonumber\\
&\qquad\qquad\qquad E_{q_{\balpha_{k}}(\bh_{k,i}|\bx_{k,i}) q_{\bbeta}(\bz_{k,i}|\bh_{k,i})} [\log p_{\btheta_k,\bgamma}(\bx_{k,i}|\bz_{k,i})] \big{)} \bigg{\}}
\end{align}
Through straightforward algebraic manipulations, the first expectation term in Equation (\ref{eq:hvae-vi}) simplifies to a squared reconstruction error:
\begin{align}\label{reconst}
    E_{q_{\balpha_{k}}(\bh_{k,i}|\bx_{k,i}) q_{\btheta_{k}}(\bz_{k,i}|\bh_{k,i})} [\log p_{\bbeta,\bgamma}(\bx_{k,i}|\bz_{k,i})] = -||\bx_{k,i} -\widehat{\bx}_{k,i}||_2^2,
\end{align}
where $\widehat{\bx}_{k,i}=\bmu_{D_k}(\bmu_{\bar{D}}(\bz_{k,i};\bgamma);\btheta_k)$ represents the reconstruction of the $k$th input image for subject $i$ obtained by passing the deep latent feature $\bz_{k,i}$ through the shared decoder followed by the image-specific decoder. This term measures the fidelity of the reconstruction: smaller values of the squared error indicate that the decoder network can accurately recover the input image from the learned latent representation. The second term in Equation (\ref{eq:hvae-vi}) acts as a regularization penalty for the image-specific latent features $\bh_{k,i}$ and assumes a closed form,
\begin{align}\label{klshallow}
    \text{KL}\big{(}  q_{\balpha_{k}}( \bh_{k,i}|  \bx_{k,i}) | p(\bh_{k,i}) \big{)} =
    \frac{1}{2}\sum_{j=1}^{p}( -\log \sigma_{E_k}(\bx_{k,i};\balpha_k)^2 + \mu_{E_k,j}(\bx_{k,i};\balpha_k)^2 + \sigma_{E_k}(\bx_{k,i};\balpha_k)^2 - 1 ),
\end{align}
where $\mu_{E_k,j}(\bx_{k,i};\balpha_k)$ is the $j$th element of $\bmu_{E_k}(\bx_{k,i};\balpha_k)$. The third term performs an analogous role for the shared deep features $\bz_{k,i}$, and also assumes a closed form, 
\begin{align}\label{eq:kldeep}
    \text{KL}\big{(} q_{\bbeta}( \bz_{k,i}| \bh_{k,i}) | p(\bz_{k,i}) \big{)}   = 
    \frac{1}{2}\sum_{j=1}^q( -\log \sigma_{\bar{E}}(\bh_{k,i};\bbeta)^2 + \mu_{\bar{E},j}(\bh_{k,i};\bbeta)^2 + \sigma_{\bar{E}}(\bh_{k,i};\bbeta)^2 - 1 ),
\end{align}
where $\mu_{\bar{E},j}(\bh_{k,i};\bbeta)$ corresponds to the $j$th element of $\bmu_{\bar{E}}(\bh_{k,i};\bbeta)$. Equation (\ref{reconst}), (\ref{klshallow}) and (\ref{eq:kldeep}) together leads to the reconstruction error of 
\begin{align}
&\mathcal{L}_{reconstruction}(\{\balpha_k,\btheta_k:k=1,..,K\},\bbeta,\bgamma)= \sum_{i=1}^n\sum_{k=1}^K\Bigg[||\bx_{k,i} -\widehat{\bx}_{k,i}||_2^2+\nonumber\\
 &\quad\frac{1}{2}\sum_{j=1}^q( -\log \sigma_{\bar{E}}(\bh_{k,i};\bbeta)^2 + \mu_{\bar{E},j}(\bh_{k,i};\bbeta)^2 + \sigma_{\bar{E}}(\bh_{k,i};\bbeta)^2 - 1 )+\nonumber\\
 &\quad \frac{1}{2}\sum_{j=1}^{p}( -\log \sigma_{E_k}(\bx_{k,i};\balpha_k)^2 + \mu_{E_k,j}(\bx_{k,i};\balpha_k)^2 + \sigma_{E_k}(\bx_{k,i};\balpha_k)^2 - 1 )\Bigg].
\end{align}

\noindent\textbf{Prediction loss:} For the supervised prediction task, the latent feature vector $\bg_i$ is passed through the DNN-based predictor as shown in Equation~\eqref{predDNNn} to produce $\widehat{y}_i$. The prediction loss is
\begin{align}
\mathcal{L}_{prediction}(\bdelta)=\sum_{i=1}^n (y_i-\widehat{y}_i)^2,    
\end{align}
which corresponds to the negative log-likelihood under a Gaussian predictive model with isotropic variance.
This supervised term not only improves predictive accuracy but also acts as an inductive bias on the encoders—pushing them to extract features that are predictive of the target while still being useful for reconstruction.

\section{Stochastic Gradient Descent for Weight and Bias Parameters}

To minimize the loss in \eqref{eq:total_loss}, 
the encoder parameters $\balpha_k$ and $\bbeta$, decoder parameters of $\bgamma$, and $\btheta_k$, and prediction parameters of $\bdelta$ are updated through stochastic gradient descent (SGD) algorithm. These parameters control how \texttt{InVA} maps data into the latent space (via $\balpha_k$ and $\bbeta$) and reconstructs it back into the original data space (via $\bgamma$ and $\btheta_k$), as well as how to  generate the final response prediction (via $\bdelta$).

\noindent\textbf{Gradient Updates for the Encoder.} For the encoder, the gradients with respect to $\balpha_k$ and $\bbeta$ are:
 \begin{align}
        \nabla_{\balpha_k,\bbeta} \mathcal{L}_{reconstrunction} &= \nabla_{\balpha_k,\bbeta}\sum_{i=1}^n\mathcal{L}(\{\balpha_k,\btheta_k:k=1,..,K\},\bbeta,\bgamma, \bx_{(i)})\nonumber\\
&= \nabla_{\balpha_k,\bbeta}\Bigg[\sum_{i=1}^n\sum_{k=1}^K \bigg{\{} \text{KL}\big{(}  q_{\balpha_{k}}(\bh_{k,i}|\bx_{k,i}) | p(\bh_{k,i}) \big{)} + \text{KL}\big{(} q_{\bbeta}(\bz_{k,i}|\bh_{k,i}) | p(\bz_{k,i})-\nonumber\\
&\qquad\qquad\qquad E_{q_{\balpha_{k,i}}(\bh_{k,i}|\bx_{k,i}) q_{\bbeta}(\bz_{k,i}|\bh_{k,i})} [\log p_{\btheta_k,\bgamma}(\bx_{k,i}|\bz_{k,i})] \big{)} \bigg{\}}\Bigg]
    \end{align}
\noindent \textbf{Gradient Updates for the Decoder.} For the decoder parameters $\bgamma$ and $\btheta_k$, the gradient update takes the form:
\begin{align}
    \nabla_{\bgamma,\btheta_k} \mathcal{L} = -\nabla_{\bgamma,\btheta_k} \sum_{i=1}^n\sum_{k=1}^K E_{q_{\balpha_k}(\bh_{k,i}|\bx_{k,i})q_{\btheta_k}(\bz_{k,i}|\bh_{k,i}}[\log p_{\bbeta,\bgamma}(\bx_{k,i}|\bz_{k,i})],
\end{align}
which focuses purely on maximizing the expected data reconstruction likelihood.

\noindent\textbf{Reparameterization trick.} Since the expectations above involve latent variables $\bh_{k,i}$ and $\bz_{k,i}$, direct gradient computation is infeasible due to their stochastic sampling. To overcome this, we use the reparameterization trick, expressing latent variables as deterministic transformations of model parameters and auxiliary noise:
\begin{align}
\bh_{k,i}&=\bmu_{E_k}(\bx_{k,i};\balpha_k)+\sigma_{E_k}(\bx_{k,i};\balpha_k)\bepsilon_h,\:\:\bepsilon_h\sim N(\bzero,\bI)\nonumber\\
\bz_{k,i}&=\bmu_{\bar{E}}(\bh_{k,i};\bbeta)+\sigma_{\bar{E}}(\bx_{k,i};\bbeta)\bepsilon_z,\:\:\bepsilon_z\sim N(\bzero,\bI).
\end{align}
This formulation ensures differentiability, allowing efficient gradient computation via backpropagation through the sampling process.

With reparameterized latent variables, the parameters are updated using SGD as follows:
\begin{align*}
        (\balpha_k^T,\bbeta^T)^T \leftarrow (\balpha_k^T,\bbeta^T)^T - \lambda \nabla_{\balpha_k,\bbeta} \mathcal{L},\:\:
         (\btheta_k^T, \bgamma^T)^T  \leftarrow (\btheta_k^T, \bgamma^T)^T - \lambda \nabla_{\btheta_k,\bgamma} \mathcal{L},
\end{align*}
where $\lambda$ is the learning rate.

Additionally, The gradient of the prediction loss $\mathcal{L}_{prediction}(\bdelta)$ with respect to each parameter in $\bdelta$ is computed by backpropagation through the predictor network as 
\begin{align}
   \mathbf{W}^{(y)}_{l} \leftarrow \mathbf{W}^{(y)}_{l} - \lambda \, \frac{\partial \mathcal{L}_{prediction}(\bdelta)}{\partial \mathbf{W}^{(y)}_{l}}, 
\quad
\mathbf{b}^{(y)}_{l} \leftarrow \mathbf{b}^{(y)}_l - \lambda \, \frac{\partial \mathcal{L}_{prediction}(\bdelta)}{\partial \mathbf{b}^{(y)}_{l}}, 
\end{align}
where $\lambda$ denotes the learning rate. This ensures that the latent embeddings learned by the encoder–decoder framework are also informative for response prediction, tightly coupling representation learning with supervised objectives.

\section{Simulation Studies}

We generate simulated 3D input and output images to assess the image prediction accuracy of our \texttt{InVA} in comparison to other baseline methods. To evaluate the models, we employ the out-of-sample mean squared prediction error (MSPE) between the output images and the predicted  images as our comparison metric, with a smaller MSPE indicating better prediction performance. The specifics of the simulation settings are provided in Section~\ref{sec:simu-set}.

\subsection{Simulation Settings}\label{sec:simu-set}

\textbf{Simulation Design:} For the $i$-th subject, where $i=1,...,n$, we generate two input images, $\bx_{1,i}$ and $\bx_{2,i}$, with each being a 3-way tensor having dimensions $d\times d \times d$, comprising of the input images having $J_1=J_2=J=d^3$ cells. Although the proposed \texttt{InVA} framework is not designed to explicitly exploit the tensor structure of these images, we adopt this representation in order to facilitate fair comparison with competing methods designed for tensor-valued regression. The cell intensities of both input images are independently simulated from a standard normal distribution: $x_{1,i}(\bj), x_{2,i}(\bj)\stackrel{i.i.d.}{\sim} N(0,1)$, where $\bj=(j_1,j_2,j_3)$ indexes cell locations of the three-dimensional grid. Each cell of the outcome image $\by_i$ is constructed according to a nonlinear polynomial regression model:
\begin{align}\label{eq:output_const}
y_i(\bj)=\sum_{o=1}^O\sum_{k=1}^2 \beta_{o,k}(\bj)x_{k,i}(\bj)^o+\epsilon_i(\bj),  
\end{align}
where $\epsilon_i(\bj)\stackrel{i.i.d.}{\sim} N(0,\sigma^2)$ represents cell-specific noise.
The simulation of the output image in Equation~\eqref{eq:output_const} implies that the dimension of the output image is same as the dimension of the input image, i.e., $m=J=d^3$ in our simulations. Here, $O$ controls the polynomial order and thereby the complexity of the relationship between inputs and the outcome. 

\noindent\textbf{Simulation Scenarios:} To comprehensively evaluate performance, we vary several key factors,
\begin{itemize}
\item \textbf{Polynomial order:} $O=1,2,3$, allowing us to examine increasing levels of nonlinearity in the outcome generation process.
\item \textbf{Noise level / Signal-to-noise ratio (SNR):} Controlled through
$\sigma\in\{0.1,0.3,0.5\}$. Smaller $\sigma$ values correspond to higher SNR and easier prediction tasks, while larger values yield noisier data.
\item \textbf{Sample size and image dimension:} We consider two combinations of $(n,d)$,
\begin{itemize}
    \item $(n,d)=(100,2):$ representing small-sample, low-dimensional settings.
    \item $(n,d)=(800,3):$ mimicking larger-scale, moderate-dimensional regimes. This latter case closely resembles the scale of multi-modal neuroimaging data examined in Section~\ref{sec:real_data}.
\end{itemize}
\end{itemize}
\textbf{Test Data:} For each setting, we generate additional test samples equal to 20\% of the training set size, following the same generative process. This allows for systematic evaluation of predictive accuracy and uncertainty quantification under matched simulation conditions.



\noindent \textbf{Baseline Competitors:} We benchmark the proposed \texttt{InVA} framework against several state-of-the-art alternatives to evaluate its performance and highlight the benefits of integrating multiple imaging modalities. First, we compare \texttt{InVA} with a standard Variational Autoencoder (VAE) model. For this, we separately use either $\bx_1=\{\bx_{1,i}:i=1,..,n\}$ or $\bx_2=\{\bx_{2,i}:i=1,..,n\}$ as input, in order to assess the potential loss of predictive power when information from one input image is ignored. These baselines are denoted as VAE($\bx_1$) and VAE($\bx_2$), respectively. In addition, we compare \texttt{InVA} with three widely used image-on-image regression approaches: (i) \emph{Bayesian Varying Coefficient Model (Var-Coef)} \citep{guhaniyogi2022distributed}, which flexibly models spatially varying relationships while allowing for nonlinear effects, (ii) \emph{Bayesian Additive Regression Trees (BART)} \citep{chipman1998bayesian}, a nonparametric method capable of capturing highly nonlinear and interaction effects between imaging inputs and outcomes, and (iii) \emph{Tensor Regression (TensorReg)} \citep{lock2018tensor}, which directly exploits the tensor structure of image data by formulating a regression model between tensor-valued inputs and outcomes. Both Var-Coef and BART are designed to handle nonlinear input–output associations, as InVA does. By contrast, TensorReg is specifically tailored to tensor-valued inputs and outcomes, but only allows linear parametric relationship between outcome and input images.

\subsection{Outcome Image Prediction Performance}
In the setting with a relatively small sample size ($n=100$) and low-dimensional images ($d=2$), we observe a general deterioration in the performance of all competing methods as the data-generating process becomes more challenging. Specifically, higher noise variance ($\sigma$) or increased complexity in the outcome–input relationship—reflected by a higher-order polynomial (i.e., higher value of $O$) governing the outcome image leads to larger prediction errors across the board.

As shown in Table~\ref{tb:mse}, \texttt{InVA} consistently delivers the lowest mean squared prediction error (MSPE) over the test datasets across almost all levels of noise and polynomial orders, outperforming TensorReg in particular. This improvement underscores \texttt{InVA}’s ability to capture intricate nonlinear dependencies between outcome and input images, where TensorReg—despite leveraging the tensor structure—falls short, perhaps due to not accounting for the nonlinear dependence between input and outcome images. While VAE($\bx_1$), VAE($\bx_2$), and BART are also designed to capture nonlinear associations, their predictive accuracy is substantially weaker than that of \texttt{InVA}. Importantly, the fact that \texttt{InVA} achieves markedly lower MSPE compared to VAE baselines demonstrates the tangible benefit of borrowing strength across multiple imaging modalities, rather than modeling each input in isolation. An interesting exception arises with the Bayesian varying coefficient model (Var-Coef). When the outcome image is generated with a simple linear dependence ($O=1$), Var-Coef performs exceptionally well because the fitted model coincides with the true data-generating mechanism. However, when the polynomial order is increased to $O=2$ or $O=3$, \texttt{InVA} comprehensively surpasses Var-Coef. This highlights a key advantage of our approach: it remains robust and adaptive in situations where the underlying relationship is complex and unknown, conditions under which simpler models may fail.

\linespread{1.2}
\begin{table*}[!t]
\setlength\tabcolsep{4.5pt} 
\caption{Mean squared prediction error (MSPE) comparison between our \texttt{InVA} and the variational autoencoder model (VAE) using only one input image, Bayesian varying coefficient model (Var-Coef), Bayesian additive regression trees (BART), and tensor regression (TensorReg) at $n=100$ and $d=2$. Across different signal-to-noise ratios, our \texttt{InVA} outperforms baseline methods when the polynomial capturing the effect of input images in the truth is of higher order ($O=2,3$), and is one of the best methods when $O=1$.}
\label{tb:mse}
\begin{center}
\begin{scriptsize}
\begin{tabular}{cccccccccccccc}
\toprule \toprule
\multirow{2}{*}{Method} & \multirow{2}{*}{Data}& & \multicolumn{3}{c}{$O=1$} &  & \multicolumn{3}{c}{$O=2$} & & \multicolumn{3}{c}{$O=3$} \\
\cline{4-6}  \cline{8-10}   \cline{12-14}
~ & ~ & & $\sigma=0.1$ & $\sigma=0.3$ & $\sigma=0.5$&  & $\sigma=0.1$ & $\sigma=0.3$ & $\sigma=0.5$ & & $\sigma=0.1$ & $\sigma=0.3$ & $\sigma=0.5$ \\
 \cline{1-2} \cline{4-6} \cline{8-10} \cline{12-14}
VAE & $\bx_1$ & & 2.80 & 2.88 & 3.01 & & 15.12 & 15.20 & 15.35 & & 134.31 & 134.56 & 135.32 \\
VAE & $\bx_2$ & & 2.78 & 2.89 & 2.98 & & 15.14 & 15.18 & 15.32 & & 134.29 & 134.59 & 135.29 \\
Var-Coef & $\bx_1$ \& $\bx_2$ & & \textbf{0.01}  & \textbf{0.01}  & \textbf{0.25} & & 22.86 & 22.95 & 23.16 &  & 61.27 & 61.17 & 61.15 \\
BART & $\bx_1$ \& $\bx_2$ & & 0.36  & 0.43  & 0.51 & & 5.18 & 5.30 & 5.39 &  & 130.63 & 130.67 & 130.84 \\
TensorReg & $\bx_1$ \& $\bx_2$ & & 3.98  & 4.08  & 4.21 & & 15.89 & 15.95 & 16.05 &  & 192.74 & 192.82 & 192.96 \\
\texttt{InVA} & $\bx_1$ \& $\bx_2$ & & 0.27 & 0.41 & 0.69 & & \textbf{2.75} & \textbf{2.86} & \textbf{3.10} & & \textbf{54.92} & \textbf{55.21} & \textbf{55.39} \\
\bottomrule
\bottomrule
\end{tabular}
\end{scriptsize}
\end{center}
\end{table*}

\linespread{1.2}
\begin{table*}[!t]
\setlength\tabcolsep{4.5pt} 
\caption{Mean squared prediction error (MSPE) comparison between our \texttt{InVA} and the variational autoencoder model (VAE), Bayesian additive regression trees (BART), and tensor regression (TensorReg) at $n=800$ and $d=3$. Across different signal-to-noise ratios and polynomial orders, our \texttt{InVA} outperforms baseline methods.}
\label{tb:mse-2}
\begin{center}
\begin{scriptsize}
\begin{tabular}{cccccccccccccc}
\toprule \toprule
\multirow{2}{*}{Method} & \multirow{2}{*}{Data}& & \multicolumn{3}{c}{$O=1$} &  & \multicolumn{3}{c}{$O=2$} & & \multicolumn{3}{c}{$O=3$} \\
\cline{4-6}  \cline{8-10}   \cline{12-14}
~ & ~ & & $\sigma=0.1$ & $\sigma=0.3$ & $\sigma=0.5$&  & $\sigma=0.1$ & $\sigma=0.3$ & $\sigma=0.5$ & & $\sigma=0.1$ & $\sigma=0.3$ & $\sigma=0.5$ \\
 \cline{1-2} \cline{4-6} \cline{8-10} \cline{12-14}
VAE & $\bx_1$ & & 4.10 & 4.19 & 4.34 & & 11.31 & 11.44 & 11.52 & & 60.23 & 60.41 & 60.75 \\
VAE & $\bx_2$ & & 4.12 & 4.21 & 4.32 & & 11.35 & 11.43 & 11.56 & & 60.26 & 60.37 & 60.72 \\
BART & $\bx_1$ \& $\bx_2$ & & 2.13 & 2.24 & 2.31 & & 14.77 & 14.85 & 14.94 & & 93.61 & 93.85 & 94.12 \\
TensorReg & $\bx_1$ \& $\bx_2$ & & 5.58 & 5.65 & 5.77 & & 21.82 & 21.93 & 22.01 & & 78.20 & 78.45 & 78.71 \\
\texttt{InVA} & $\bx_1$ \& $\bx_2$ & & \textbf{0.49} & \textbf{0.62} & \textbf{0.82} & & \textbf{5.72} & \textbf{5.78} & \textbf{6.17} & & \textbf{36.52} & \textbf{36.75} & \textbf{36.82} \\
\bottomrule
\bottomrule
\end{tabular}
\end{scriptsize}
\end{center}
\end{table*}

\linespread{1.2}
\begin{table*}[!t]
\setlength\tabcolsep{4.5pt} 
\caption{Ablation studies: Mean squared prediction error comparison between our \texttt{InVA} and our \texttt{InVA} without shared components (\texttt{InVA} w/o Shd) and our \texttt{InVA} without input image-specific components (\texttt{InVA} w/o IS) at $n=100$ and $d=2$. Across different signal-to-noise ratios and polynomial orders, our \texttt{InVA} outperforms \texttt{InVA} w/o Shd and \texttt{InVA} w/o IS, demonstrating the importance of both the input image-specific and shared components in our \texttt{InVA}.}
\label{tb:mse-abla}
\begin{center}
\begin{scriptsize}
\begin{tabular}{cccccccccccccc}
\toprule \toprule
\multirow{2}{*}{Method} & \multirow{2}{*}{Data}& & \multicolumn{3}{c}{$\text{order}=1$} &  & \multicolumn{3}{c}{$\text{order}=2$} & & \multicolumn{3}{c}{$\text{order}=3$} \\
\cline{4-6}  \cline{8-10}   \cline{12-14}
~ & ~ & & $\sigma=0.1$ & $\sigma=0.3$ & $\sigma=0.5$&  & $\sigma=0.1$ & $\sigma=0.3$ & $\sigma=0.5$ & & $\sigma=0.1$ & $\sigma=0.3$ & $\sigma=0.5$ \\
 \cline{1-2} \cline{4-6} \cline{8-10} \cline{12-14}
\texttt{InVA} w/o Shd& $\bx_1$ \& $\bx_2$ & & 3.42 & 3.49 & 3.58 & & 11.48 & 11.54 & 11.62 & & 152.10 & 152.26 & 152.45 \\
\texttt{InVA} w/o IS & $\bx_1$ \& $\bx_2$ & & 1.48 & 1.55 & 1.61 & & 5.78 & 5.85 & 5.95 & & 101.63 & 101.84 & 102.08 \\
\texttt{InVA} & $\bx_1$ \& $\bx_2$ & & \textbf{0.27} & \textbf{0.41} & \textbf{0.69} & & \textbf{2.75} & \textbf{2.86} & \textbf{3.10} & & \textbf{54.92} & \textbf{55.21} & \textbf{55.39} \\
\bottomrule
\bottomrule
\end{tabular}
\end{scriptsize}
\end{center}
\end{table*}

In the case of $n=800$ and $d=3$, \texttt{InVA} continues to outperform the baseline competitors (refer to Table~\ref{tb:mse-2}). Var-Coef is not included as a baseline due to computational challenges with $n=800$. Similar to Table~\ref{tb:mse}, Table~\ref{tb:mse-2} demonstrates a decline in performance with increasing noise variance and the order of the true data-generating polynomial. Importantly, both tables establish significantly superior performance when information is suitably borrowed from the two input images in predicting the outcome image.


\subsection{Ablation Studies}

To further assess the contribution of different architectural components, we conduct ablation studies on our proposed \texttt{InVA}. Specifically, we evaluate two variants: (i) \texttt{InVA} w/o Shd, where shared components are removed, and (ii) \texttt{InVA} w/o IS, where input image–specific components are excluded. In the \texttt{InVA} w/o Shd variant, each input image is equipped with its own encoder and decoder, and predictions are obtained by averaging over modalities. Importantly, this version does not include a shared encoder–decoder pair, thereby eliminating the mechanism for explicitly capturing information common across input images. Conversely, in the \texttt{InVA} w/o IS variant, we pool all modalities and train only a shared encoder and decoder, without including input-specific encoders and decoders. This setup allows the model to exploit shared structure across images but ignores image-specific variations, which may contain important predictive signals.

We retain mean squared prediction error (MSPE) as the evaluation metric and present the results in Table~\ref{tb:mse-abla}. Across all experimental settings—varying both signal-to-noise ratios and polynomial orders—our full \texttt{InVA} consistently achieves lower MSPE than either ablated variant. The ablation results highlight the necessity of a hybrid design that balances common and image-specific structures, thereby enabling robust harmonization and improved predictive accuracy in multi-modal neuroimaging analysis.

\subsection{Computation Time}

\begin{figure*}[!t]
    \centering
    \includegraphics[width = 0.5\textwidth]{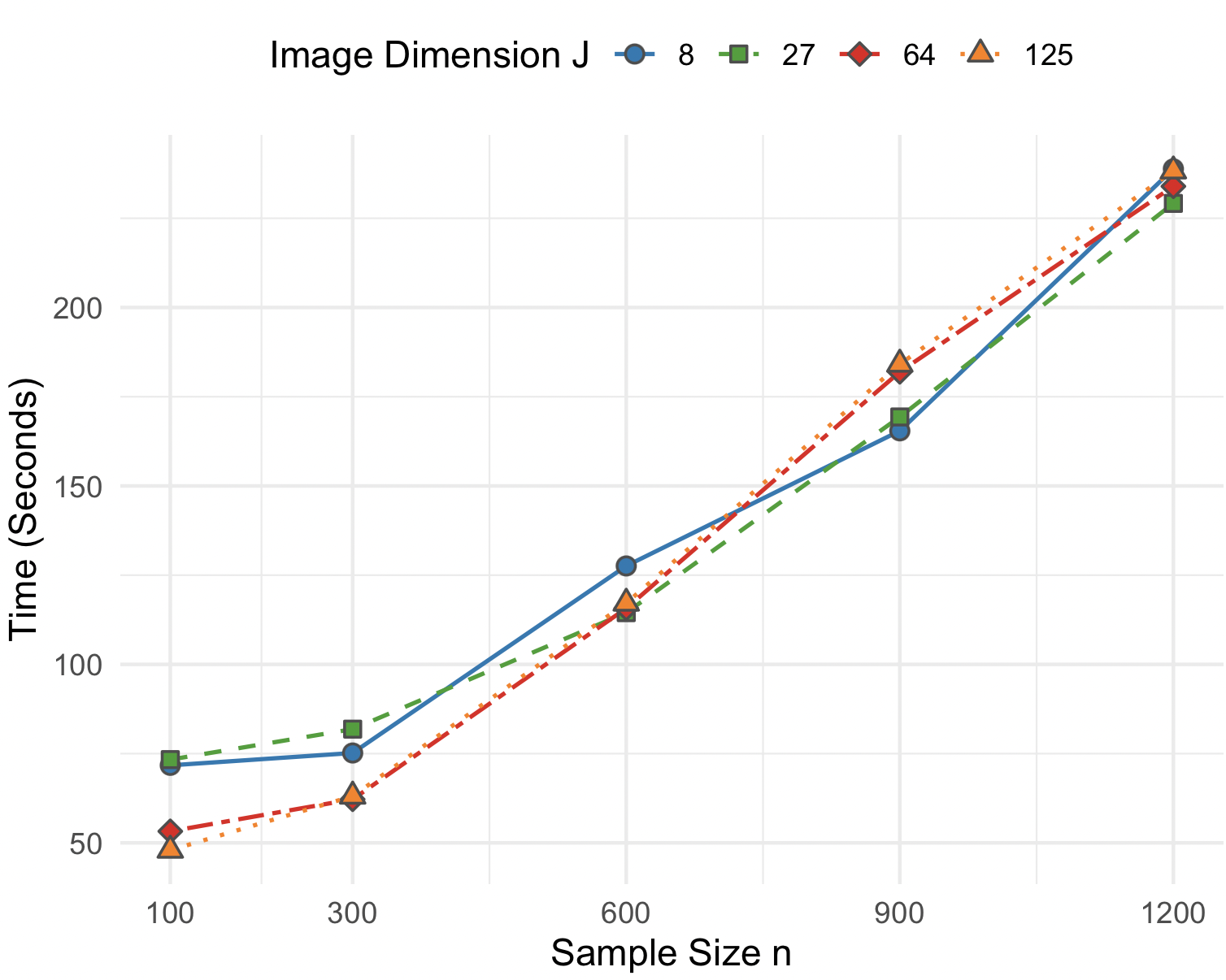}
    \caption{Computation time versus sample size $n$ across image dimension $J$.
Lines show the mean total wall–clock time per run (100 epochs; batch size 64; fixed architecture; no early stopping).
Shades encode different values of $J$.}\label{fig:compute_analysis}
\end{figure*}

We fixed the layer widths and number of training epochs (without early stopping) and varied the sample size $n=100,300,600,900,1200$, while letting the dimension of the input tensor image governed by different choices of $d=2,3,4,5$. This corresponds to input images of 
$J=d^3=8,27,64,125$ cells, respectively. As illustrated in Figure~\ref{fig:compute_analysis}, the training time scales almost linearly with $n$, reflecting the fact that the number of batches per epoch grows proportionally with sample size while all other training parameters are held constant.

By contrast, the effect of input dimension $J$ on training time is mild and not strictly monotone. This behavior is consistent with the design of our \texttt{InVA} architecture: while the initial input projection (encoder) and final output projections (decoder/predictor) scale with $J$, the bulk of computation occurs in hidden layers of fixed width, which are invariant to $J$. For very small inputs (e.g., $J=8$, corresponding to $d=2$), per-batch overhead and suboptimal kernel utilization can dominate, occasionally making training slower than for larger $J$ despite the reduced outcome size.

Overall, these results demonstrate that training time is governed primarily by sample size rather than input dimension, with the dependence on $J$ being secondary and largely implementation-specific. Importantly, this highlights the scalability of our approach and its practical suitability for large-scale neuroimaging studies, where rapid training and efficient computation are critical.

\section{Multi-modal Neuroimaging Data Analysis}\label{sec:real_data}



We further apply our \texttt{InVA} approach in the study of multi-modal neuroimaging data. Data used in the preparation of this article were obtained from the Alzheimer’s Disease Neuroimaging Initiative (ADNI) database (adni.loni.usc.edu)\footnote{Data used in preparation of this article were obtained from the Alzheimer’s Disease Neuroimaging Initiative (ADNI) database (adni.loni.usc.edu). As such, the investigators within the ADNI contributed to the design and implementation of ADNI and/or provided data but did not participate in analysis or writing of this report. A complete listing of ADNI investigators can be found at: \href{http://adni.loni.usc.edu/wp-content/uploads/how_to_apply/ADNI_Acknowledgement_List.pdf.}{http://adni.loni.usc.edu/-/}.}. 
The primary goal of ADNI has been to test whether serial MRI, PET, other biological markers, and clinical and neuropsychological assessment can be combined to measure the progression of AD. Specifically, we consider the baseline visit for participants in the ADNI 1, GO, and 2 cohorts. The goal of this analysis is to model molecular A$\beta$ PET images as a function of MRI images of cortical thickness and volume. To do so, PET and MRI images were registered to a common template space and segmented into 40 regions of interest (ROI) via the Desikan-Killiany cortical atlas \citep{Desikan2006} using standard ADNI pipelines as described in \citet{tadpole2019}. Measurements of A$\beta$ deposition were characterized by standardized uptake value ratio (SUVR) images which detect A$\beta$ via binding of the florbetapir radiotracer. Cortical thickness and volume were extracted and measured in millimeters (mm) and $\text{mm}^3$ using FreeSurfer \citep{Fischl2012}. Complete imaging data was available for 711 subjects whose clinical status ranged from some cognitive impairment to a diagnosis of AD. The goal in this data is to predict the PET image using cortical thickness and cortical volume obtained from MRI. To assess predictive performance of the proposed method, we randomly divide the data into two parts, one part 80\% as the training set, and one part 20\% as the test set. 
To ensure robust evaluation, we conducted repeated validation in which the data were randomly split into training and test sets across multiple runs (repeated 50 times).This procedure allows us to assess not only the average predictive accuracy but also the stability of each method. In our comparisons, all baseline competitors mentioned in Section 3.1 are compared with \texttt{InVA}, excluding TensorReg and Var-Coef. Var-Coef is computationally demanding for the size of the dataset, and TensorReg is not applicable to the dataset since the input and output images are not tensors in the real data, unlike in our simulation settings. 

\subsection{Prediction Comparison with Repeated Training-Test Split}\label{sec:train-test-pred}
The average runtime over 50 repetitions for the proposed \texttt{InVA} approach is 7.29 seconds. In comparison, VAEs trained solely on cortical thickness or cortical volume required 6.29 seconds, indicating that the inclusion of an additional deep layer in \texttt{InVA} does not substantially increase computational burden. By contrast, Bayesian Additive Regression Trees (BART), even when implemented with optimized code in the \texttt{BART} package in \texttt{R}, required an average of 12.24 seconds, nearly double the runtime of \texttt{InVA}.

\begin{figure}[ht]
\centering\includegraphics[scale=0.45]{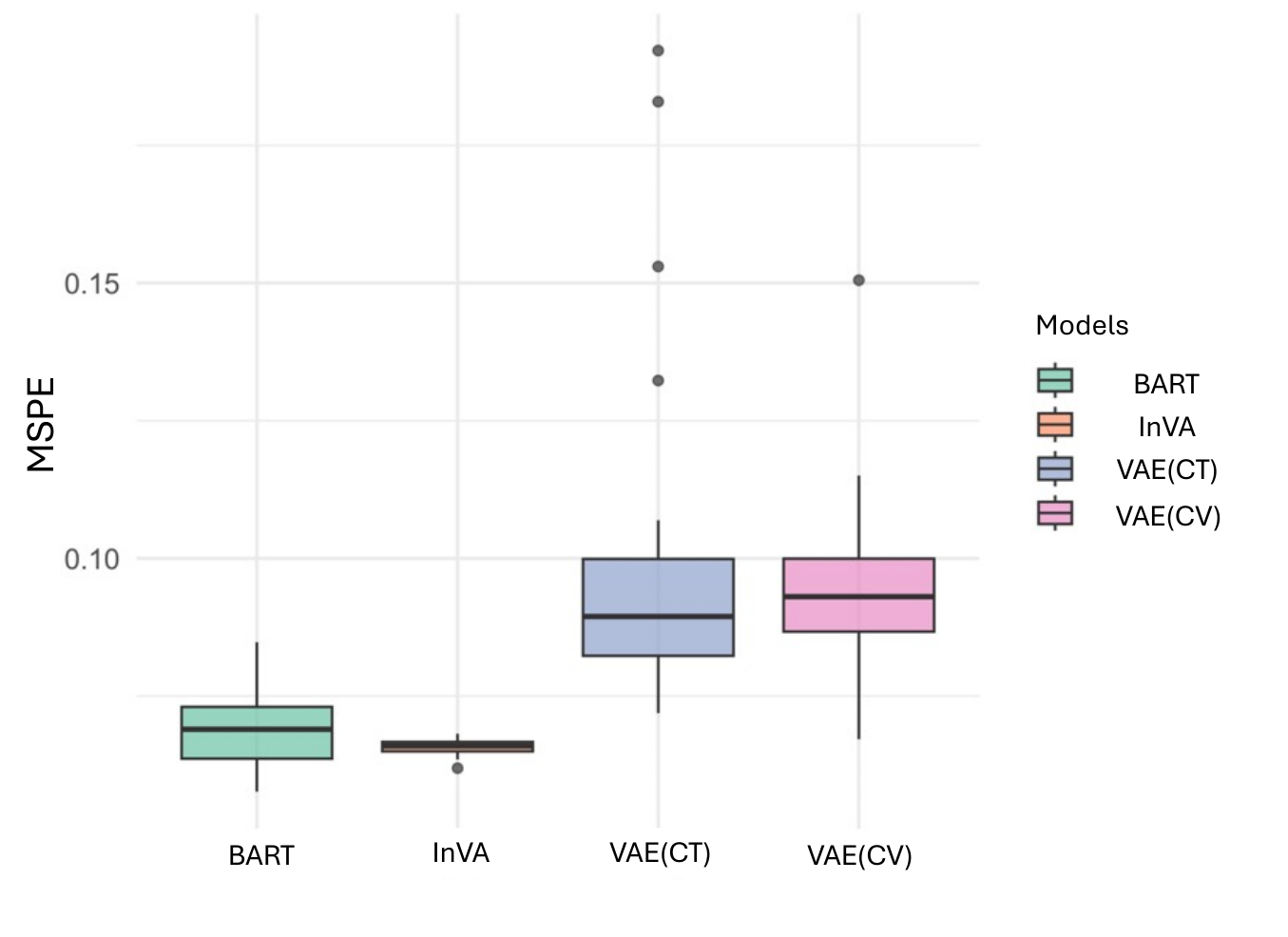}
    \caption{The figure presents boxplots of mean squared prediction errors (MSPE) across 50 training–test repetitions on the real dataset for all competing methods. Competitors include the proposed \texttt{InVA}, a standard VAE trained on either cortical volume or cortical thickness, and BART. The results demonstrate that \texttt{InVA} achieves both a lower average MSPE and substantially reduced variability across repetitions, highlighting its superior predictive accuracy and greater stability compared to alternative approaches.}
\label{fig:results_repeated}
\end{figure}

As summarized in Figure~\ref{fig:results_repeated}, \texttt{InVA} consistently outperforms competing methods in terms of predictive accuracy, achieving the lowest test mean squared prediction error (MSPE) across 50 independent repetitions. Moreover, the variability of MSPE values over repetitions under \texttt{InVA} is notably reduced, reflecting its stability and robustness. In contrast, BART achieves slightly higher predictive error but exhibits greater variability across repetitions. The VAEs, whether based on cortical thickness or cortical volume, perform substantially worse, yielding higher MSPE and much greater variability, underscoring their limited predictive utility in this setting. Taken together, these findings highlight that \texttt{InVA} achieves superior predictive performance while maintaining computational efficiency. Its training times are on par with, or even shorter than, widely used alternatives, demonstrating that the methodological advances in \texttt{InVA} translate into practical gains in both accuracy and efficiency.

\subsection{Predictive Inference on ROIs}

To further examine predictive performance at the regional level, we select a representative training–test split from the 50 repetitions and evaluate model performance across different ROIs. This representative split yields MSPE for the competing models as reported in Table~\ref{tb:mse-4}, confirming that the chosen split is consistent with overall trends observed across all repetitions. Our primary focus here is on ROI-level prediction of PET imaging outcomes using cortical volume and cortical thickness as inputs under the proposed \texttt{InVA} framework.
\linespread{1.2}
\begin{table}[!t]
\setlength\tabcolsep{4.5pt} 
\caption{Mean squared prediction error (MSPE) for predicting PET images from cortical volume and cortical thickness is compared across \texttt{InVA}, the variational autoencoder (VAE), and Bayesian additive regression trees (BART) for one representative training–test split of the multi-modal neuroimaging data. The proposed \texttt{InVA} achieves the smallest MSPE, with values consistent with the overall trend reported in Section~\ref{sec:train-test-pred}, confirming that the selected split is representative.}
\label{tb:mse-4}
\begin{center}
\begin{scriptsize}
\begin{tabular}{ccc}
\toprule \toprule
Method & Data& MSPE \\
\hline
VAE & Cortical Thickness & 0.0803 \\
VAE & Cortical Volume & 0.1092 \\
BART & Cortical Thickness \& Volume & 0.0667\\
\texttt{InVA} & Cortical Thickness \& Volume & \textbf{0.0660}\\
\bottomrule
\bottomrule
\end{tabular}
\end{scriptsize}
\end{center}
\end{table}

\begin{figure}[ht]
\centering\includegraphics[scale=0.38]{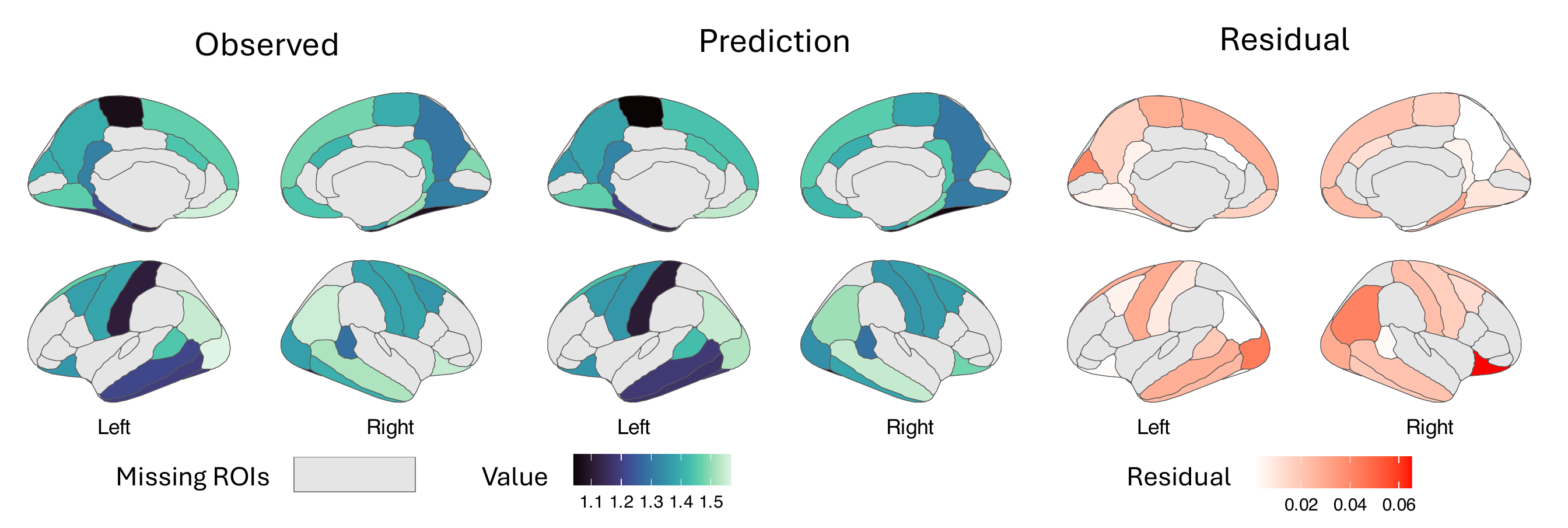}
    \caption{Observed and predicted PET images for ROI-wise average across all subjects, along with their residuals. Gray regions correspond to missing ROIs not defined in the Desikan–Killiany atlas. The observed and predicted PET image show strong similarity, suggesting that the observed PET response is accurately reconstructed using the estimated PET response, while the residuals highlight ROI-specific variations in error.}
\label{fig:results}
\end{figure}

In Figure~\ref{fig:results}, the average PET response (averaged over all subjects) is observed alongside the estimated average PET response along with their difference, capturing error in the estimated mean, illustrating the accurate reconstruction of the observed PET response. The error varies by ROI giving us insight into which regions of the brain admit better recovery of the PET signal from cortical thickness and volume. Note that the gray regions correspond to missing ROIs, reflecting the fact that the Desikan–Killiany atlas defines only 40 cortical regions; regions not included in this atlas are shown in gray. The lowest errors are observed in the \emph{caudal anterior cingulate} and the \emph{precuneus} which have been observed in prior studies to exhibit amyloid driven changes to cortical structures \citep{Becker2011Amyloid}. Thus, their strong association here echoes prior findings. The highest errors are observed in the \emph{lateral orbitofrontal} region which was found to lack significant differences in cortical thickness between amyloid positive and negative groups in prior work \citep{Fan2018} suggesting a lack of information in cortical structures that can be used to predict amyloid levels. Together, these observations demonstrate that \texttt{InVA} is able to recapitulate prior observed patterns of association between cortical structures and amyloid deposition.


\section{Conclusion and Discussions}
We introduce a novel integrative variational autoencoder approach designed to leverage information from multiple imaging inputs, allowing for the development of a nonlinear relationship between input images and an image output. While there is an existing literature on hierarchical VAE approaches, to our knowledge, \texttt{InVA} is the first hierarchical VAE that exploits individual and shared information in multiple imaging inputs to predict an imaging outcome. The proposed approach also allows model-free image-on-image regression capturing complex non-linear dependence between input and outcome images. Empirical results from simulation studies demonstrate the superior performance of our proposed approach compared to existing image-on-image regression methods, particularly in drawing predictive inferences on the outcome image. This approach holds transformative potential in the field of multi-modal neuroimaging, especially in accurately predicting costly tau-PET images using more affordable imaging modalities for the study of neurodegenerative diseases, such as Alzheimer's.

Despite the harmonization of multi-modal neuroimaging data modeling, this article does not comprehensively explore our approach for a gamut of other multi-modal perspective, such as text data, video data, and audio data~\citep{jabeen2023review, xu2023multimodal}. We plan to explore this issue in a future article. Additionally, it is intuitive that our integrative variational autoencoder can be combined with existing uni-modal VAEs to equip each encoder and decoder component with a more expressive architecture. Finding the optimal combination and design remains to be explored, and this will be a future research direction.


\section{Acknowledgements}

Rajarshi Guhaniyogi acknowledges funding from National Science Foundation Grant DMS-2210672 and National Institute Of Neurological Disorders And Stroke of the National
Institutes of Health under Award Number R01NS131604. Aaron Scheffler acknowledges funding from National Science Foundation Grant DMS-2210206 and the National Institute Of Neurological Disorders And Stroke of the National
Institutes of Health under Award Number R01NS131604. The content is solely the responsibility of the authors and does not necessarily represent the official views of the National Science Foundation or the National Institutes of Health.

Data collection and sharing for this project was funded by the Alzheimer's Disease Neuroimaging Initiative
(ADNI) (National Institutes of Health Grant U01 AG024904) and DOD ADNI (Department of Defense award
number W81XWH-12-2-0012). ADNI is funded by the National Institute on Aging, the National Institute of
Biomedical Imaging and Bioengineering, and through generous contributions from the following: AbbVie,
Alzheimer’s Association; Alzheimer’s Drug Discovery Foundation; Araclon Biotech; BioClinica, Inc.; Biogen;
Bristol-Myers Squibb Company; CereSpir, Inc.; Cogstate; Eisai Inc.; Elan Pharmaceuticals, Inc.; Eli Lilly and
Company; EuroImmun; F. Hoffmann-La Roche Ltd and its affiliated company Genentech, Inc.; Fujirebio; GE
Healthcare; IXICO Ltd.; Janssen Alzheimer Immunotherapy Research \& Development, LLC.; Johnson \&
Johnson Pharmaceutical Research \& Development LLC.; Lumosity; Lundbeck; Merck \& Co., Inc.; Meso
Scale Diagnostics, LLC.; NeuroRx Research; Neurotrack Technologies; Novartis Pharmaceuticals
Corporation; Pfizer Inc.; Piramal Imaging; Servier; Takeda Pharmaceutical Company; and Transition Therapeutics. The Canadian Institutes of Health Research is providing funds to support ADNI clinical sites in Canada. Private sector contributions are facilitated by the Foundation for the National Institutes of Health
(www.fnih.org). The grantee organization is the Northern California Institute for Research and Education, and the study is coordinated by the Alzheimer’s Therapeutic Research Institute at the University of Southern California. ADNI data are disseminated by the Laboratory for Neuro Imaging at the University of Southern California.


\bibliography{example_paper}
\bibliographystyle{abbrvnat}

\end{document}